\documentclass[12pt]{article}

\input epsf.tex
\def\DESepsf(#1 width #2){\epsfxsize=#2 \epsfbox{#1}}

\begin{document}

\begin{titlepage}
\begin{flushright}
OSU-HEP-00-07\\
BA-00-66\\
December 2000\\
\end{flushright}
\vskip 2cm
\begin{center}
{\Large\bf The Scaling of Lepton Dipole Moments\\[0.15in]
with Lepton Mass}
\vskip 1cm
{\normalsize\bf
K.S.\ Babu$\,{}^1$ S.M. Barr$^2$ and Ilja Dorsner$^2$ }
\vskip 0.5cm
{\it ${}^1\,$Department of Physics, Oklahoma State University,\\
Stillwater, OK~~74078 \\ [0.15truecm]
${}^2\,$Bartol Research Institute, University of Delaware, \\Newark,
DE~~19716\\[0.1truecm]
}

\end{center}
\vskip .5cm

\begin{abstract}

The dipole moments of the leptons and quarks are matrices in flavor space,
which can potentially reveal as much about the flavor structure of the theory as
do the mass matrices. The off-diagonal elements of the dipole matrices
lead to flavor-changing decays such as $\mu \longrightarrow e \gamma$,
while the imaginary parts of the diagonal elements give rise to electric dipole
moments. We analyze the scaling of the leptonic dipole moments with the
lepton masses in theories beyond the standard model.  While in many models the dipole
moments scale roughly as lepton mass, it is shown that simple models exist
in which the dipoles scale as the cube of the mass or in other ways. An explicit
example with cubic scaling is presented, which is motivated on independent grounds
from large angle neutrino oscillation data.  Our results have great significance
for the observability of the electric dipole moments
$d_e$, $d_{\mu}$, $d_{\tau}$, and the rare decays $\mu \longrightarrow e \gamma$,
and $\tau \longrightarrow \mu \gamma$ and will be tested in several forthcoming
experiments.

\end{abstract}
\end{titlepage}

\newpage

\section{Introduction and summary of results}
Most of the parameters of the standard model are masses and mixings of
the quarks and leptons. These masses and mixings exhibit patterns
that suggest to many theorists that there exist underlying
``flavor" or ``family" symmetries. Over the years many models of fermion
masses and flavor symmetry
have been proposed, some of them quite elegant. However, it is
impossible without further experimental facts to
decide among the various theoretical ideas. A determination of neutrino
masses and mixings will certainly be enormously important for
solving the ``flavor problem", but probably even more experimental facts
will be required.

Another window on the flavor problem could be provided by the
dipole moments of the quarks and leptons, since not only
the fermion masses but the dipoles as well
should reflect the underlying flavor symmetries
of the theory. In fact, potentially there could be
as much information about flavor in the dipole moments
as in the masses and mixings. Given $N$ fermions $f_i$ having the same
Standard Model quantum numbers, their anomalous dipole moments, like their
masses, form an $N \times N$
complex matrix, which we shall denote as $D_{ij}$. This matrix is the
coefficient of the effective operator $f_{iL}^T C \sigma_{\mu \nu}
F^{\mu \nu} f^c_{jL}$.
Whereas in the ``physical basis" of the fermions
the mass matrix becomes real and diagonal, the anomalous dipole moment matrix
would be expected to remain complex and, in general, non-diagonal.

The imaginary part of the diagonal elements of the
anomalous dipole moment matrix corresponds to the
electric dipole moments of the particles: $Im(D_{ii}) \equiv d_i$.
The off-diagonal elements of the anomalous
dipole moment matrix are the ``transition dipole moments",
which can be either magnetic or electric: $Re(D_{ij}) \equiv \mu_{ij}$, and
$Im(D_{ij}) \equiv d_{ij}$.
The transition
dipole moments lead to decays such as
$\mu \longrightarrow e \gamma$ \cite{raredecays}.

In this paper we shall focus on the dipole moments of the
charged leptons. These are quite interesting from the experimental
point of view as it is likely that there will soon be great
improvements in the limits on several of these quantities.
The present limit on the electric dipole moment (edm)
of the electron, coming from
atomic experiments, is $4.3 \times 10^{-27}$ ecm \cite{thallium}. A
recently approved experiment on PbO expects to push this down to the
level of $10^{-29}$ ecm in three to five years, and hopes
eventually to reach even the level of $10^{-31}$ ecm \cite{demille}.
The current limit on the muon edm is $1.1 \times
10^{-18}$ ecm \cite{dmu}, but there have been  recent proposals to push this
to $10^{-24}$ ecm \cite{dmuprop}.
The present limit on the branching ratio of $\mu
\longrightarrow
e \gamma$ is $1.2 \times 10^{-11}$ \cite{meg}. There is a proposal
to push this to the level of $10^{-14}$ at PSI \cite{megprop}.
The forthcoming MECO experiment \cite{MECO} will improve the limit
on $\mu-e$ conversion in nuclei to the level of $10^{-16}$ in branching
ratio, which is directly
related to the $\mu \rightarrow e \gamma$ decay rate.
The Brookhaven experiment E821 is currently in the process of improving
the precision on muon $g-2$ to the level of (few) $\times 10^{-10}$ \cite{bnl}.
The current limit on
$\tau \longrightarrow \mu \gamma$ is
$1.1 \times 10^{-6}$
\cite{tmg}.
This limit will also be
substantially improved, at least to the $10^{-7}$ level, and perhaps
to $10^{-8}$ \cite{tmgprop}. The current limit
on the edm of the tau lepton is
$3.1 \times 10^{-16}$ ecm \cite{dtau}, which could be improved significantly.

In this paper we shall show that there are reasonable and well-motivated
models in which several of these leptonic dipole moments,
specifically $D_{\mu e}$, $d_{\mu}$, and $D_{\tau \mu}$,
may be within reach of experiment.
If this were so, then the pattern of the leptonic edm matrix would
be extremely revealing of the underlying flavor structure of the theory.
It is also quite possible, of course, that no lepton edm or transition
dipole moment is
large enough to be seen. Certainly this is the case in the Standard
Model, where these quantities are highly suppressed. However, if there
is non-trivial flavor structure at scales between a few hundred
GeV and a few hundred TeV, then some of these moments may be observable.
If there is no new physics all the way to about a thousand TeV, then
there is very little chance for seeing these edm's or the rare leptonic decays
in the forthcoming experiments.  However, new physics such as supersymmetry
is anticipated not much above a few TeV, so
chances for seeing these effects are great, as we will argue.

In many models there is a close relationship between the
anomalous dipole moment matrices and the mass matrices of the fermions.
Both kinds of matrix connect the left-handed fermions to the
right-handed fermions. Moreover, at the diagrammatic level
the Feynman diagrams that contribute to dipole moments
and masses are related. In particular,
removing the external photon from a dipole moment diagram produces
a mass diagram, and adding a photon to a mass diagram of one or more loops
produces a dipole diagram. In fact, in many models the anomalous
dipole moments of the quarks and leptons are roughly proportional
to their masses.

In the minimal supersymmetric
Standard Model (MSSM), for instance, there are well-known one-loop
contributions to lepton and quark edms coming from gaugino loops
\cite{susyedm}.
In those diagrams a lepton (or quark) emits a virtual gaugino to
become a virtual slepton (or squark) and then reabsorbs
the gaugino, coupling in the process to an external photon line.
Since a dipole moment operator involves a chirality flip of the fermion,
there must be an insertion either of the mass of the external
fermion or of the ``left-right" mass of the corresponding
virtual sfermion. However, the left-right
sfermion masses, $m^2_{LR}$, are
proportional to the masses of the corresponding fermions in the MSSM. Thus
the edms of the quarks and leptons come out being approximately
proportional to their masses.

Proportionality of edms and masses also occurs in multi-Higgs models.
In models with more than one Higgs doublet there can be CP violation
in the Higgs propagators that can lead to measurable edms \cite{weinberg}.
The dominant contribution to the edms of the lighter quarks and leptons
come from two-loop diagrams that involve
one power of the Higgs Yukawa couplings \cite{barrzee}.
These contributions are therefore approximately
proportional to the mass of the quark or lepton. (There are also
one-loop diagrams, but these are cubic in the Higgs Yukawa
couplings and are therefore negligible for the lighter families.
These one--loop contributions may however become significant
for the muon edm for some range of parameters, especially large
$\tan\beta$ \cite{barger}.)

This feature that the anomalous dipole moments
scale roughly as the first power of
the fermion mass is characteristic of many models. If it does
in fact hold in Nature, then it has several important implications.
First, there would be a limit on the muon edm coming from the
present limit on the electron edm:
$d_{\mu} \sim (m_{\mu}/m_e) d_e \leq 8.8 \times 10^{-25}$ ecm.
This is at the margin of what could be seen by the proposed
experiments. Second, one would conclude that
$d_{\tau}$ and $D_{\tau \mu}$ are unobservably small.
For example, one would have
$d_{\tau} \sim (m_{\tau}/m_e) d_e \leq 1.5 \times 10^{-23}$ ecm,
which is $0.5 \times 10^{-7}$ of the present limit and is
well beyond the reach of currently planned experiments.

However, the anomalous dipole moments
of the leptons and quarks do not necessarily
scale as the first power of the masses. In this paper we shall
describe models in which they
scale as other powers, in particular as the
cube or the square. The idea that electric dipole moments might scale
as the cube of the mass
is not new. As already noted, the one-loop contributions to the
fermion edms in multi-Higgs models do tend to scale with the
cube of the mass, and these were for a time naturally thought
to be the leading contributions. However, it was then
pointed out \cite{barrzee}
that two-loop diagrams which scale as the first power
of mass dominate for the lighter families.

We present an explicit model in Sec. 2 and 3 where the dipoles obey
a cubic scaling.  The model is motivated on independent grounds, viz.,
an explanation of solar and atmospheric neutrino oscillation data in
terms of large angle neutrino oscillations.
In this model we find that the anomalous dipole moment matrix scales roughly as
follows:

\begin{equation}
D_{ij} \sim d_\tau \left(
\begin{array}{ccc}  (m_e/m_\tau)^3 + c {\alpha \over 4 \pi}(m_e/m_\tau) & -~~~
& - \\ & & \\
(m_em_{\mu}^2/m_\tau^3) &
(m_{\mu}/m_\tau)^3 ~~~& - \\ & & \\ (m_{e}/m_\tau)^2 &
(m_{\mu} m_{\tau}) ~~~& 1
\end{array} \right).
\end{equation}

\noindent
Here the dashes indicate elements that are negligibly small.
One sees that the diagonal elements --- and therefore the edms of the
particles ---
scale as the mass cubed, while those elements
below the diagonal are suppressed relative to the diagonal elements
by factors that go as the mass ratios (coming from mixing angles).

In Eq. (1), the dipole moment of the electron has a term that is denoted
as $c {\alpha \over 4 \pi}(m_e/m_\tau)d_\tau$.  Since the electron mass is
so small, one has to take into account two loop diagrams for the dipole
moments which may have a linear scaling with the lepton mass \cite{barrzee}.
Typically, we find that the coefficient $c$ is of order unity.  With $c =
{\cal O}(1)$,the two loop diagram with linear scaling will win over the
one--loop diagram with cubic scaling. But
this is significant only for the electron dipole moment.
It is also possible that these two--loop diagrams are suppressed in some
models, in which case $c \simeq 0$, although in the explicit models that we
have constructed, this turns out not to be the case.

Consider the case of $c \simeq 0$ (i.e., the two--loop diagrams being suppressed).
The resulting pattern is interesting from the point of view of several
kinds of experiments. If we suppose that the transition dipole
$D_{\mu e}$ is just at the current limit coming from
$B(\mu \longrightarrow e \gamma) \leq 1.2 \times 10^{-11}$, namely
$\left| D_{\mu e} \right|
= 4.5 \times 10^{-27}$ ecm, then the pattern in Eq. (1) with $c=0$ would
imply the following values of other moments:
$d_e \sim 10^{-31}$ ecm, $d_{\mu} \sim 10^{-24}$ ecm, $d_{\tau}
\sim 5 \times 10^{-21}$ ecm, and $\left| D_{\tau \mu} \right|
\sim 3 \times 10^{-22}$
ecm (which corresponds to $B(\tau \longrightarrow \mu \gamma)
\sim 10^{-4}$). It should be emphasized that these are very rough estimates,
which could well be off by an order of magnitude in the dipole moment
or two orders of magnitude in a rare decay rate. Making allowances for this,
one can say that if the pattern in Eq. (1) is correct then:
(a) $d_e$ is probably too small to
be seen in the near future, though experiments on molecules like
PbO have a chance of seeing it eventually.
(And even if not seen, the failure to see it together with the observation
of $d_{\mu}$ would confirm a more-than-linear scaling of the edms
with mass.) (b) $d_{\mu}$
should be at or near the level which recently proposed experiments can
reach. And (c) $\tau \longrightarrow \mu \gamma$ should be very close
to the present limit.

In the more realistic case of $c = {\cal O}(1)$, the above estimates will
still hold, except for the electron edm.  $d_e$ should be also observable,
at the level of $1 \times 10^{-27}$ ecm, if the decay $\mu \longrightarrow
e \gamma$ is near the present experimental limit.

In section 4, we shall describe, in less detail, a model that
illustrates quadratic scaling of the anomalous dipole moments with mass.
In that model one finds

\begin{equation}
D_{ij} \sim d_\tau
\left( \begin{array}{ccc} (m_e/m_\tau)^2 & (m_e m_\mu/m_\tau^2) & (m_e/m_\tau) \\
& & \\ (m_e m_\mu/m_\tau^2) & (m_{\mu}/m_\tau)^2 & (m_{\mu}/m_{\tau}) \\
& & \\ (m_e/m_\tau) & (m_{\mu}/m_{\tau}) & 1
\end{array} \right).
\end{equation}

In this case the two--loop contribution to $d_e$ which would scale linearly
with the electron mass is at best of the same
order numerically as the one--loop correction
and is not shown in Eq. (2).  If $\mu \longrightarrow e \gamma$ is
saturated at the present limit, this pattern would lead to the following
predictions:  (a) $d_e \simeq 2 \times 10^{-29}$ ecm, (b) $d_\mu \simeq
1 \times 10^{-24}$ e-cm, (c) $d_\tau \simeq 3 \times 10^{-22}$ ecm,
(d) $|D_{\tau \mu}| \simeq 2 \times 10^{-23}$ ecm (corresponding to
$B(\tau \longrightarrow \mu \gamma) \sim 4 \times 10^{-7}$).
Again, we see that these predictions can be well tested directly in the
near future.

\section{Flavor symmetry and the dipole moments}

Before describing a specific model that leads to the pattern shown in
Eq. (1), we will describe in more general terms how different kinds
of scaling of the lepton edms can arise in simple models based
on abelian flavor symmetry.

Many models of fermion masses explain the interfamily mass hierarchy by
appealing to spontaneously broken flavor symmetry. The idea is
that certain elements of the mass matrix would vanish if some
abelian symmetry or symmetries
were exact, and are therefore suppressed by powers of
the expectation values that violate those symmetries. The fields whose
expectation values violate flavor symmetry are often called ``flavons".
If $F$ denotes a flavon field, then different elements of the
mass matrix could be suppressed by different powers of $\epsilon \equiv
\langle F \rangle/M_F$, where $M_F$ is a scale characterizing the
underlying flavor physics. The absolute scale of $M_F$ and
$\langle F \rangle$ could be anything from just above the weak
scale up to the Planck scale. All that matters for explaining
the fermion mass hierarchy is that their ratio $\epsilon$ be smaller
than, but of order, unity. Some models have several flavon fields, and
have the fermion masses depending on several small parameters
$\epsilon_i$.

In a typical model of the kind we have been describing the fermion
masses arise from tree-level diagrams similar to that shown in
Fig. 1(a) \cite{fn}. One of the
boson lines attached to the fermion line
is a Higgs doublet whose vacuum expectation value breaks
$SU(2)_L \times U(1)_Y$. The others are flavon fields, whose
vacuum expectation values give rise to the requisite powers of
the small parameter(s) $\epsilon_i$. It is evident that
if two of the boson lines are tied together to make a loop, and
an external photon line is attached, the resulting diagram
(Fig. 1(b)) gives
a non-zero dipole moment to the fermion. If the flavor scale is near
the unification scale or Planck scale, the resulting dipole moment
is negligible. However, if the flavor scale is near
the weak scale, a significant dipole moment can result. One sees,
moreover, that the dipole matrix will have a close relationship to the
mass matrix.

Consider the following simple toy model as an illustration.
This model will be generalized in the next section to a fully realistic model that also
explains large neutrino mixing angles as suggested by recent
data from solar and atmospheric neutrino experiments.  The toy model
is over--simplified in that it has vanishing inter--generational mixings.
Suppose one has the ordinary three families of leptons ($\ell_i$) and
three families also of singly-charged vectorlike leptons ($X_i$):

\begin{equation}
\left( \begin{array}{c}
\nu \\ \ell^- \end{array} \right)_{iL} \equiv L_i, \;\;\; \ell^+_{iL}, \;\;\;
X^-_{iL}, \;\;\; X^+_{iL}.
\end{equation}

\noindent
Suppose the following Yukawa couplings and mass terms:

\begin{equation}
{\cal L}_{Yuk} =
\sum_{i} f_i (L_i \ell^+_{iL}) H^{\dag} +
\sum_{i,K} f_i^K(X^-_{iL} X^+_{iL}) h_K +
\sum_{i} f_{i}^S (X_{iL}^- \ell_{iL}^+) S_i.
\end{equation}

\noindent
This structure can be enforced by a simple flavor symmetry.
The field $H$ is the usual Higgs doublet, while $S_i$ and $h_K$
are singlets. Call $f_i \langle H^0 \rangle \equiv \hat{m}_i$,
$\Sigma_K f_i^K \langle h_K \rangle \equiv \langle F_i \rangle$,
and $f_i^S \langle S_i \rangle \equiv M_i \equiv \epsilon_i^{-1}
\langle F_i \rangle$,
where it is assumed that
$\epsilon_i < 1$ and
$m_i \ll M_i$, $\langle F_i \rangle$. Then there is a mass matrix
of the form

\begin{equation}
(\ell^-_{iL}, X^-_{iL}) \left( \begin{array}{cc}
\hat{m}_i & 0 \\ M_i & \langle F_i \rangle \end{array} \right)
\left( \begin{array}{c} \ell^+_{iL} \\ X^+_{iL} \end{array} \right).
\end{equation}

\noindent
This matrix has three large eigenvalues that are of order $M_i$,
and three small eigenvalues that are of order $\hat{m}_i$. The three
light eigenstates are to be identified as the physical $e$, $\mu$
and $\tau$.
Neglecting terms of order $(\hat{m}_i/M_i)$, the large entries in
the mass matrix are
diagonalized by a change of basis of the positively charged leptons:
$\ell^{+ \prime}_{iL} = \sin \theta_i \ell^+_{iL}
- \cos \theta_i X^+_{iL}$, and $X^{+ \prime}_{iL} =
\cos \theta_i \ell^+_{iL} + \sin \theta_i X^+_{iL}$,
where $\tan \theta_i = \epsilon_i$.
(We have implicitly chosen a phase convention wherein the entries in
the mass matrices have all been made real.  The couplings of the
leptons to the scalar fields will however remain complex in this
basis.)  This would give effective mass terms for the light three
fermions of the form $m_i (\ell^-_{iL}
\ell^{+ \prime}_{iL})$, with

\begin{equation}
m_i = \frac{\langle F_i \rangle}{M_i} \hat{m}_i = \hat{m}_i \epsilon_i,
\end{equation}

\noindent
where we have used $\sin \theta_i
\cong \tan \theta_i = \epsilon_i$. We suppose that the parameters
$\hat{m}_i$ are all of the same order and that the fermion
mass hierarchy comes from a hierarchy in the $\epsilon_i$:
$\epsilon_1 \ll \epsilon_2 \ll \epsilon_3$.

Diagrammatically, one has at lowest order in the
$\epsilon_i$ the tree-level diagram in Fig. 2(a). This would
give an effective mass term for the light fermions of
$(f_i \langle H \rangle) (f_i^S \langle S_i \rangle)^{-1}$\newline
$(\Sigma_K f_i^K \langle h_K \rangle) = \hat{m}_i M_i^{-1} \langle F_i \rangle
= \hat{m}_i \epsilon_i$,
which of course agrees with the result of diagonalizing
the matrix.

If an external photon line is added to the diagram in Fig. 2(a)
and the $H$ and $h_K$ boson lines are tied together through
the quartic term $\lambda_K H^{\dag} H h_K^* h_K$, which must certainly
be allowed to exist by symmetry, the diagram shown in Fig. 2(b)
results. The contribution to the dipole moment of the $i^{th}$ light lepton
($\ell_1 = e$, $\ell_2 = \mu$, and $\ell_3 = \tau$) will be of the form

\begin{equation}
d_i = \frac{1}{16 \pi^2} \sum_{K} (e \lambda_K f_i f_i^K)
\langle H \rangle \langle h_K \rangle M_i
I(m^2_H, m^2_{h_K}, M^2_i),
\end{equation}

\noindent
where $I$ is the function with dimensions of $M^{-4}$
resulting from doing the momentum integral,
given by $I(r_1 M, r_2 M, M) = \frac{1}{2} M^{-4}(r_1 - r_2)^{-1}
([1 - r_1^2(1 + \ln r_1^2)]/(1 - r_1)^3 -
(r_1 \longrightarrow r_2))$.
Let us suppose that the largest mass in the
loop is that of the heavy virtual fermion (which Eq. (5) shows
to be approximately $M_i$).
Then the integral is given by
$I  \cong (M_i)^{-4} (\frac{3}{2} + \frac{5}{2}(m_H^2 + m_{h_K}^2)/M_i^2))$,
Recalling that $\Sigma_K f_i^K \langle h_K \rangle
= \langle F_i \rangle$, define $\Sigma_K \lambda_K f_i^K
\langle h_K \rangle \equiv \lambda_i \langle F_i \rangle$.
If one neglects the $K$ dependence in the momentum integral
$I$, the dipole moment of the $i^{th}$ light lepton can be written as

\begin{equation}
d_i \cong  \frac{3e}{32 \pi^2} \lambda_i
\frac{\hat{m}_i \langle F_i \rangle}{M_i^3}
= \frac{3e}{32 \pi^2} \lambda_i \frac{m_i}{M_i^2}~.
\end{equation}

\noindent
It is technically natural to assume that the CP-violating
phase of $\lambda_i$ is of order one. It should be noted that there will
be such a phase in the physical basis of the leptons,
in general, if there is more than one type of boson $h_K$.
The point is that the mass of the $i^{th}$ light lepton, which is a real number
in the physical basis, is proportional to $\Sigma_K f_i^K
\langle h_K \rangle$, whereas the dipole of that lepton depends on
$\Sigma \lambda_K f_i^K \langle h_K \rangle$, which has no reason
to be real in the same basis. Thus the electric dipole moment of
the $i^{th}$ lepton is of the same order as the result in Eq. (8)

One can see that this edm can be quite large for reasonable values of
the parameters. For example, if $M_2 = 3$ TeV and $\lambda_2 = 1$,
then the edm of the muon would be of order $3 \times 10^{-24}$ ecm.
The scaling of the edms with mass depends upon how the
family mass hierarchy is assumed to arise. This hierarchy
is given by $m_i/m_j \sim \epsilon_i/\epsilon_j = (\langle F_i \rangle/
\langle F_j \rangle) (M_i/M_j)^{-1}$. Thus, it can come from
(a) a hierarchy in the $\langle F_i \rangle$, with all the $M_i$
being comparable, (b) a hierarchy in the $M_i$,
with all the $\langle F_i \rangle$ being comparable, or
(c) a combination of these.
If (a) is the case, then Eq. (8)
would imply that the edms of the leptons scale approximately {\it linearly}
in lepton mass. However, if (b) is the case, then the edms scale with
the cube of the masses, since $m_i/m_j \sim M_j/M_i$ and thus
$d_i/d_j \sim (m_i/m_j)(M_j/M_i)^2 \sim (m_i/m_j)^3$.

Since the one--loop dipoles scale as the cube of the lepton masses if the
hierarchy is in $M_i$, one should
examine whther two--loop diagrams induce larger contributions, especially
for $d_e$.  We find that this is indeed the case, as in multi--Higgs models.
Diagonalizing the mass matrix of Eq. (5) to lowest order in $\epsilon_i$
and $\hat{m}_i/M_i$ will lead to the following Yukawa couplings of the light
fermions to the scalars:

\begin{equation}
{\cal L}_{\rm Yuk} = \ell_{iL}^- \ell_{iL}^{+\prime}\epsilon_i f_i H +
\ell_{iL}^- \ell_{iL}^{+\prime} {\hat{m}_i \over M_i} f_i^K h_k + H.C.
\end{equation}

\noindent In the physical basis, $\epsilon_i \hat{m_i}$ are real, but $f_i^K$ are
complex in general.  Thus, an edm of the electron can be induced through a
two--loop diagram with $h_K$ propagating.  The dominant contribution to $d_e$
will arise when $h_K$ decays into a real and a virtual photon.  This happens,
for example, by a diagram in which $h_K$ emits  an $X_{iL}^+ X_{iL}^-$
pair which closes by emitting two photons,
or one where $h_K$ becomes an $H$ (through quartic scalar couplings),
which then decays to two photons through
a top quark (and other) loop. All these diagrams, while they are suppressed
by an additional loop factor, scale linearly with the lepton mass and can be
dominant over the one--loop diagram.  For $m_{h_K} \simeq 300$ GeV, $d_e$ turns
out to be of order $4 \times 10^{-27}$ ecm \cite{barrzee}, if the relevant
phases are all of order one.

\section{A realistic model with mixing of charged lepton families}

The simple example just discussed is perhaps too simple in that
it does not have any mixing among the families. It is true that
the leptonic mixing seen in neutrino oscillations might arise
from the neutrino sector rather than the charged lepton sector.
However, the CKM mixing of the quarks and the likelihood that quarks
and charged leptons are unified in some way suggest otherwise.
In this section we present a unifiable model of lepton mass
in which the charged lepton mass
matrix is off-diagonal in the flavor basis. This model is actually
a realization of an idea proposed several years ago
to explain both the fermion mass hierarchy and the largeness of
neutrino mixing angles \cite{babubarr}. The model is thus well-motivated on
grounds having nothing to do with dipole moments. But it
naturally leads to the pattern given in Eq. (1).

The model has the following leptons and scalar fields:

\begin{equation}
\begin{array}{ccccccccc}
{\rm field:} & \;\; & \ell_i^- & \ell_i^+ & X_i^- & X_i^+ & H & S_i & h_K \\
& & & & & & & & \\
F \; {\rm charge:} & & 0 & 0 & q_i & -q_i & 0 & -q_i & 0 \\
& & & & & & & &
\end{array}
\end{equation}

\noindent
where $F$ is an abelian flavor symmetry. This symmetry allows the
following Yukawa terms for the leptons:

\begin{equation}
{\cal L}_{Yuk} = \sum_{i,j} f_{ij} ( \ell_i^- \ell_j^+) H^0
+ \sum_i f_i^K ( X_i^- X_i^+) h_K +
\sum_{i,j} f_{ij}^S ( X_i^- \ell_j^+) S_i.
\end{equation}

\noindent
The crucial difference between this model and the toy model
discussed in the last section is that here the Yukawa coupling
$f_{ij}$ is a matrix. The coupling $f_{ij}^S$ is also a matrix,
but this is of less significance, as we shall see.
It will simplify the discussion without affecting the qualitative
conclusions if we assume that
$f_{ij}^S$ is diagonal: $f_{ij}^S = f_i^S \delta_{ij}$.
This assumption will be relaxed later.

Let us define the matrix $\hat{m}_{ij} = f_{ij} \langle H^0 \rangle$.
There will be a $6 \times 6$ mass matrix of the charged leptons,
which has the same form as in the toy model (cf. Eq. (5)) except that
$\hat{m}_{ij}$ is here non-diagonal:

\begin{equation}
(\ell_{iL}^-, X_{iL}^-) \left( \begin{array}{cc}
\hat{m}_{ij} & 0 \\ M_i \delta_{ij} & \langle F_i \rangle \delta_{ij}
\end{array} \right) \left( \begin{array}{c}
\ell_{jL}^+ \\ X_{jL}^+ \end{array} \right).
\end{equation}

\noindent
The quantities $M_i$ and $\langle F_i \rangle$
appearing here are defined in the same way as in the toy
model, and $\epsilon_i = \langle F_i \rangle/ M_i$.
After diagonalizing the large, i.e. $O(M_i)$,
elements in the mass matrix, the light leptons (namely
$\ell^-_i$ and $\ell^{+ \prime}_j \cong X^+_j - \epsilon_j \ell^+_j$)
have an effective
mass matrix given by

\begin{equation}
m_{ij} = \hat{m}_{ij}
\frac{\langle F_j \rangle}{M_j} = \hat{m}_{ij} \epsilon_j.
\end{equation}

\noindent
As in the toy model of the last section, we assume that all the $f_{ij}$
and hence all the $\hat{m}_{ij}$ are of the same order, and that the
fermion mass hierarchy comes from the $\epsilon_i$:
$\epsilon_1 \ll \epsilon_2
\ll \epsilon_3$.
From Eq. (13) it is obvious that
the $3 \times 3$ light lepton mass matrix has a
column form. That is, the first, second and third columns of
the matrix have elements that are, respectively, small, medium, and large.
That means that in diagonalizing it, the rotations that are done from
the left (i.e. of the $\ell_{iL}^-$) are of order one, whereas those
done from the right (i.e. of the $\ell_{iL}^{+ \prime}$) are suppressed by
the hierarchy factors $\epsilon_j/\epsilon_k \sim m_j/m_k$, $j < k$.
This means that there would be large leptonic mixing angles, as
is observed in atmospheric neutrino oscillations, and perhaps also
in solar neutrino oscillations. If one were to embed this model in
$SU(5)$ then the mass matrix of the down quarks $d$, $s$, and $b$
would have a similar structure, except transposed (since $SU(5)$
relates $M_L$ to $M_D^T$, where $M_L$ and $M_D$ are the mass matrices of the
charged leptons and down quarks). As the down quark mass matrix would thus
have a {\it row}
structure instead of a column structure, it would be the mixing angles
of the $d_{iL}$ that would be of order $m_i/m_j$,
while those of the $d^c_{iL}$ would be of order one.
Thus, such a model would explain the fact that at least certain of
the leptonic (MNS) mixings are large while the quark (CKM) mixings are
small \cite{barrdorsner}.

Returning to the leptons, it is easy to see that their dipole matrix
also has a column form. In fact, the same calculation that led to
Eq. (8) gives

\begin{equation}
D_{ij} \cong \frac{3e}{32 \pi^2} \hat{m}_{ij}
\frac{\lambda_j \langle F_j \rangle}{M_j^3}
= \frac{3e}{32 \pi^2} m_{ij} \frac{\lambda_j}{M_j^2}.
\end{equation}

\noindent
Both the dipole matrix and the mass matrix of the light leptons
have a column form, and the corresponding columns of the two matrices
are in fact proportional to each other. However, the hierarchies among the
columns are different in the two matrices. (Compare Eqs. (13) and (14).)
The case we are
most interested in is where the hierarchy in the $\epsilon_i$
parameters comes from a hierarchy in the $M_i$, with the
$\langle F_i \rangle$ all being of the same order. Then the
columns of $m_{ij}$ are in the ratio $\epsilon_1:\epsilon_2:\epsilon_3$,
while the columns of $D_{ij}$ are roughly in the ratios
$\epsilon_1^3 : \epsilon_2^3 : \epsilon_3^3$. This shows that
$d_e : d_{\mu} : d_{\tau} = m_e^3 : m_{\mu}^3 : m_{\tau}^3$.

To find the magnitudes of the off-diagonal elements of $D_{ij}$
is simple. The first thing to be noticed is that because of the
fact that the columns of $m_{ij}$ and $D_{ij}$ are proportional
to each other, the same unitary transformation acting from the left will
make zero all the elements above the diagonal in both matrices.
This is a crucial fact, since otherwise there would be large
off-diagonal elements in $D_{ij}$. For instance, in the matrix
in Eq. (14) the (12) element is of the same order as the (22) element,
which leads to the danger that $d_{\mu} \sim \left| D_{e \mu} \right|$.
That would
mean that the limit on $\mu \longrightarrow e \gamma$ would put
a severe constraint on the muon edm. As it is, the elements
above the diagonal can be made zero simultaneously in both matrices,

To complete the diagonalization of $m_{ij}$
requires a unitary transformation
from the right. Because there is a strong hierarchy among the
columns of $m_{ij}$ this unitary transformation involves
rotations $U_{ij}$, $i<j$, that are of order
$\epsilon_i/\epsilon_j \ll 1$. This will not eliminate the
elements of $D_{ij}$ that are below the diagonal, since the hierarchy
among the columns of $D_{ij}$ is different. Rather, the result
of this unitary transformation will be to make $D_{ji}$, $i<j$,
be of order $(\epsilon_i/\epsilon_j) D_{jj}$. This gives the form
shown in Eq. (1).

An interesting prediction that arises from the nearly triangular
form in Eq. (1),
aside from the ones already discussed in the Introduction, is
that in the decay $\mu^- \longrightarrow e^- \gamma$ the electron
will be almost purely right-handed. A similar remark applies for
the decay $\tau^- \longrightarrow \mu^- \gamma$.

Up to this point we have oversimplified the analysis of this model
by assuming that the Yukawa matrix $f_{ij}^S$ appearing in Eq. (11)
is diagonal. The symmetries do not require it to be diagonal, so this is
an unnatural assumption. However, eliminating
this assumption makes no qualitative
difference to the form of the dipole matrix.
Suppose that $f_{ij}^S$ is allowed to take a general form. Then
the $6 \times 6$ mass matrix of the leptons is

\begin{equation}
( \ell^-_{iL}, X^-_{iL}) \left( \begin{array}{cc}
\hat{m}_{ij} & 0 \\
M_{ij} & \langle F_i \rangle \delta_{ij}
\end{array} \right) \left( \begin{array}{c}
\ell^+_{jL} \\ X^+_{jL} \end{array} \right),
\end{equation}

\noindent
where $M_{ij} \equiv f_{ij}^S \langle S_i \rangle$. As before, we will
assume that the hierarchy comes from $M_{ij} = f_{ij}^S \langle S_i \rangle$
rather than from
$\langle F_i \rangle$. Since $f_{ij}^S$ is assumed to be an arbitrary
matrix, this means that the hierarchy comes from $\langle S_i \rangle$.
In other words, $\langle S_1 \rangle \gg \langle S_2 \rangle \gg
\langle S_3 \rangle$. This means that $M_{ij}$ has a row structure,
with the first row being much larger than the second, which in turn
is much larger than the third. By rotating among the $\ell^+_{jL}$
one can go to a basis where $M_{ij}$ is triangular, with zeros above
the diagonal. This will also change the form of the Yukawa matrix
$f_{ij}$, but since that matrix is not assumed to have any special
form anyway, this makes no difference. Thus without loss of
generality, one can take $M_{ij}$ to have this triangular form.
To completely diagonalize it then requires a rotation among the
$X_i^-$ by angles that go as $\langle S_j \rangle/ \langle S_i \rangle
\sim \epsilon_i/\epsilon_j$, $i<j$. This rotation will have the effect
of making the coupling of $X^-_i$ to $X^+_j$ non-diagonal. This is
the only change from our previous analysis. However, the off-diagonal
elements thus introduced are suppressed by the small hierarchy
factors $\epsilon_i/\epsilon_j$, $i<j$. One can easily trace through
the effect of these small off-diagonal elements, and one finds that
they give rise to a mixing of the $i^{th}$ and $j^{th}$
columns of the dipole matrix
that is of order $\epsilon_i/\epsilon_j$, $i<j$. This does not
change the form of the dipole matrix given in Eq. (1).

As in the previous section, there are two--loop diagrams that are suppressed
by additional factors of $(\alpha/4 \pi) \sim 10^{-3}$, but these are still
more significant to $d_e$.  The dominant diagrams involve the exchange
of $h_K$ fields.  The coupling matrix of $h_K$ to the light fermions are
neither diagonal nor real in the physical basis for the leptons.  As in the
toy model, we would expect $d_e \sim 4 \times 10^{-27}$ ecm if the masses of
$h_K$ are of order 300 GeV.  These two--loop diagrams have negligible effects
for the muon and the tau lepton.

We have a model of mixing that is satisfactory for the leptons.
As already mentioned, it is possible to embed this structure in
an $SU(5)$ model. The simplest way to do this would be to
add to the usual three families of ${\bf 10}_i + \overline{{\bf 5}}_i$
three vectorlike pairs $\overline{{\bf 10}}_i^{\prime} +
{\bf 10}_i^{\prime}$. The Higgs $S_i$ and $h_K$ would be $SU(5)$
singlets, while $H$ would be a ${\bf 5}$. The charge assignments
in Eq. (10) and the couplings in Eq. (11) would be generalized in
the obvious way.
This would lead to a mass matrix for the
down quarks that was simply the transpose of that for the charged
leptons, just as in the minimal $SU(5)$ model.
A realistic $SU(5)$ version of this model would have to
introduce non-minimal features, such as larger Higgs multiplets
or higher-dimensional effective operators contributing to light
fermion masses. That would allow Clebsch coefficients to appear
in the mass matrices which differed for the down quarks and the
charged leptons. Nonetheless, one would expect that the down quark
mass matrix, $M_D$, would be closely related to the transpose of
the charged lepton mass matrix, $M_L^T$, even if not exactly equal to it.
As noted above, this would elegantly explain the curious fact that
all the CKM quark mixing angles have been observed to be small, while
at least some of the MNS leptonic angles are large. For $M_L$
has a column structure, while $M_D$ has a row structure.

A question arises, however, whether the model can be extended to include
quarks without running into a problem with excessive flavor changing
in the quark sector.
In particular, will one-loop diagrams involving the scalar fields
$H$, $S_i$, and $h_K$ lead to excessive $K^0-\overline{K}^0$ mixing?
The answer turns out to be no. Consider, for example, an $SU(5)$
version of this model where one takes the parameters to have the
following orders of magnitude: $f_{ij} \sim 10^{-2}$,
$f_{ij}^S \sim 1$, $f_{ij}^K \sim 1$, $\langle H \rangle \sim
300$ GeV, $\langle h_K \rangle \sim m_{h_{K}} \sim 300$ GeV --- $1$ TeV,
$\langle S_3 \rangle \sim 300$ GeV --- $1$ TeV,
$\langle S_2 \rangle \sim (\epsilon_3/\epsilon_2) \langle S_3 \rangle$,
and $\langle S_1 \rangle \sim (\epsilon_3/\epsilon_1) \langle
S_3 \rangle$, where $\epsilon_1 : \epsilon_2 : \epsilon_3 \sim
m_e : m_{\mu} : m_{\tau} \sim m_d : m_s : m_b$.
It can be shown that none of the box
diagrams involving the scalar fields $H$, $h_K$, and $S_i$,
and the new heavy vectorlike
fermions contribute excessively to either the real or
imaginary parts of $M_{12}$ in the neutral kaon system.
Some of the diagrams are suppressed by small mixings of order
$m_d/m_b$ and $m_s/m_b$ or powers thereof, or by the large masses
of the vectorlike fermions, especially $X^-_1$ and $X^-_2$, or
by the small values of the $f_{ij}$, or by combinations of all these
factors. This fact, namely that one can have in this model
one-loop dipole moments that are of interesting magnitude ---
including flavor-changing ones --- without having excessive
one-loop flavor changing in the kaon system, is non-trivial.

It might be suspected that the mixing of the light fermions with
the heavy vector fermions could lead to large flavor--changing $Z$ couplings
that can arise at the tree level.  However, this turns out to be not
the case in this model.  For example, in the lepton sector, there
is no flavor changing $Z$ coupling at the tree level in the right--handed
sector.  There are such couplings in the left--handed sector, but
they are suppressed by $(\hat{m}/M)^2$ in the amplitude, which is at most 
of order $10^{-6}$
for $f_{ij} \sim 10^{-2}$ and $M$ of order TeV.  
The resulting branching ratio for
$Z \longrightarrow
\tau \mu$ decay for example is at the level of $10^{-12}$, which is
too small to be observable, but is not much below the current limit for 
the decay $\mu \longrightarrow 3 e$.

It should be observed that the $SU(5)$ version of this model
described above could not be supersymmetrized if we wish to maintain
perturbative unification at a scale of $10^{16}$ GeV.
Having three vectorlike ${\bf 10} + \overline{{\bf 10}}$ pairs along with
supersymmetry would
destroy the perturbativity of the gauge couplings around $10^9$ GeV.

\section{Other scalings of the lepton dipole matrix}

We have seen examples of linear and cubic scaling of the electric
dipole moments with fermion mass. It is not difficult to construct
models in which the scaling is quadratic. For example, consider
a model in which there are the following leptons and Higgs fields.
Leptons: $L_i \equiv (\nu_{iL}, \ell^-_{iL})^T$,
$\ell^+_{iL}$, $n_{aL}$, $n^c_{aL}$.
Higgs: $H \equiv (H^0, H^-)^T$, $h_a \equiv (h^0_a, h^-_a)^T$,
$s^-_a$. Here $n_a$ are neutral leptons which are taken to be Dirac
particles.
It is assumed that the fields $h_a$ and $s_a$ have vanishing VEVs.
Let there be the
following couplings: $\Sigma_a m_a (n^c_{aL} n_{aL}) + \Sigma_{a,i} f_{ia}
h^{\dag}_a (L_i n_a) + \Sigma_{a,i} f'_{ia} s^-_a (\ell^+ n^c_a)
+ \Sigma_a \lambda_a h_a H (s_a^-)^*$. This form can be enforced easily with
abelian flavor symmetries that distinguish fields with different values
of the index `$a$'.

Suppose that the masses of the known leptons $\ell^{\pm}$
arise predominantly from the loops shown in Fig. 3. Let the heaviest
particles in the loops be the bosons $h_a$ and $s^-_a$, which are
assumed to have masses $\sim M_a$ that are in a hierarchy:
$M_3 \ll M_2 \ll M_1$. The largest contribution to the lepton mass
matrix $m_{ij}$ therefore comes from the diagram in Fig. 3 having $a = 3$.
This gives a contribution to $m_{ij}$ that is of rank 1. This would
generate the mass of the $\tau$ lepton. The second largest contribution
comes from the diagram with $a=2$ and also is rank 1. This would give mass
to the muon. The diagram with $a=1$, finally, gives the electron mass \cite{radiative}.
If we assume that the only hierarchy in the underlying parameters
is in the $M_a$, then it is apparent from simple power counting
that the lepton masses have a hierarchy that goes roughly as
$m_e: m_{\mu}: m_{\tau} \sim M_1^{-2}: M_2^{-2}: M_3^{-2}$, since the loop
diagram of Fig. 3 is quadratically convergent.

If one adds an external photon line to the loop, it must
couple to the boson line, since the virtual lepton $n_a$ is neutral.
The resulting diagram for the anomalous dipole moment is
quartically convergent. Thus one obtains
$d_e: d_{\mu}: d_{\tau} \sim M_1^{-4}: M_2^{-4}: M_3^{-4} \sim
m_e^2: m_{\mu}^2: m_{\tau}^2$. In fact, it is not difficult to
show that the whole anomalous dipole matrix has the scaling given in
Eq. (2).

One difference from the model of Sec. 2 and 3 is that numerically the
magnitudes of the dipole moments are enhanced here by a loop factor,
which is roughly $(1/16 \pi^2)^{-1} \sim 100$.  The observability of
the edm is thus greater in this case.

It is also possible to generate the dipole matrix of Eq. (2) by
introducing singlet charged leptons and neutral scalars instead of
neutral leptons and charged scalars.

There is an obvious generalization of this model to the quark sector.
We can introduce singlet quarks with electric charges $+2/3$ and $-1/3$.
The ordinary quarks can couple to these vector quarks
through the higgs doublets $h_a =
(h_a^0, h_a^-)$ and neutral higgs singlets $s_a$  which do not acquire
VEVs.   One--loop diagrams analogous to Fig. 3 will induce quark
masses as well as dipole moments.  If the masses of the $h_a$ fields (and/or
the $s_a$ fields) are hierarchical, the observed fermion mass hierarchy
can be reproduced, with all Yukawa couplings being of the same order.  This
might be an interesting radiative mass hierarchy model for all the
fermions, except for the top quark.

In the supersymmetric standard model, if there is non--trivial flavor structure
in the trilinear $A$ terms, analogous mass hierarchy can be induced, with
interesting dipole moments \cite{farrar}.  In this case, the diagrams will involve exchange
of the gaugino and squarks.

\section{Conclusions}

The magnetic and electric dipole moment matrices of quarks and leptons
can potentially carry a wealth of information about the underlying flavor structure.
In a large class of models we have shown that these are experimentally accessible, through
the electric dipole moments of the electron, muon and the tau lepton, and through
rare decays such as $\mu \longrightarrow e \gamma$ and $\tau \longrightarrow
\mu \gamma$.  We have addressed
the scaling of the leptonic dipole moments with the lepton mass.  While linear
scaling with the mass is quite generic, in a class of
models motivated on independent grounds
by large angle neutrino oscillation solution to the solar and
atmospheric neutrino data,
we have found interesting scalings, where the dipoles go as the cube or the
square of the lepton mass.

Dipole moments of quarks and leptons have been analyzed in the literature
in various other contexts.  Their scaling with mass in grand unified
supersymmetric models
is typically linear \cite{hall,okada}, in contrast to the
cubic scaling that we have found.  It is also possible that
significant contributions
to leptonic dipole moments arise through SUSY exchange
from interactions that are responsible for heavy
right--handed neutrino masses \cite{bdm}.  Such interactions do not lead to
any simple power law scaling.  Experiments in the near future will be
capable of telling these  theories apart.  Indeed,
very important information about the flavor structure
will be gained by the proposed improvements in the dipole moment measurements
as well as in the radiative decays of $\mu$ and $\tau$.

\section*{Acknowldgments}

The work of K.B. is supported in part by DOE Grant \# DE-FG03-98ER-41076,
\# DE-FG02-01ER4864, a grant
from the Research Corporation and by the OSU Environmental Institute.  S.B and I.D
are supported by DOE Grant \# DE-FG02-91ER-40626 A007.

\newpage

\begin{picture}(360,216)(0,0)
\thicklines
\put(36,108){\vector(1,0){36}}
\put(72,108){\line(1,0){72}}
\put(180,108){\vector(-1,0){36}}
\put(180,108){\vector(1,0){36}}
\put(216,108){\line(1,0){72}}
\put(324,108){\vector(-1,0){36}}
\thinlines
\put(108,108){\line(0,-1){36}}
\put(180,108){\line(0,-1){36}}
\put(252,108){\line(0,-1){36}}
\put(99,51){$\langle H \rangle$}
\put(171,51){$\langle F_i \rangle$}
\put(243,51){$\langle F_j \rangle$}
\put(162,9){{\bf Fig. 1(a)}}
\end{picture}
\begin{center}
A typical tree diagram for light fermion mass through flavon fields.
\end{center}

\vspace{2cm}

\begin{picture}(360,216)(0,0)
\thicklines
\put(36,144){\vector(1,0){36}}
\put(72,144){\line(1,0){72}}
\put(180,144){\vector(-1,0){36}}
\put(180,144){\vector(1,0){36}}
\put(216,144){\line(1,0){72}}
\put(324,144){\vector(-1,0){36}}
\thinlines
\put(180,144){\oval(144,144)[b]}
\put(180,144){\line(0,-1){30}}
\put(171,103){$\langle F_i \rangle$}
\put(180,72){\line(1,-1){20}}
\put(180,72){\line(-1,-1){20}}
\put(151,40){$\langle H \rangle$}
\put(191,40){$\langle F_j \rangle$}
\put(171,144){\oval(10,10)[tl]}
\put(171,154){\oval(10,10)[r]}
\put(171,164){\oval(10,10)[l]}
\put(171,174){\oval(10,10)[r]}
\put(144,162){$\gamma$}
\put(90,108){$H$}
\put(261,108){$F_j$}
\put(162,0){{\bf Fig. 1(b)}}
\end{picture}

\begin{center}
Dipole moment of the fermion induced  by folding the legs of Fig. 1(a).
\end{center}

\newpage

\begin{picture}(360,216)(0,0)
\thicklines
\put(36,108){\vector(1,0){36}}
\put(36,90){$\ell^-_i$}
\put(72,108){\line(1,0){72}}
\put(99,117){$f_i$}
\put(135,90){$\ell^+_i$}
\put(180,108){\vector(-1,0){36}}
\put(171,117){$f_i^S$}
\put(180,108){\vector(1,0){36}}
\put(216,108){\line(1,0){72}}
\put(207,90){$X^-_i$}
\put(243,117){$f_i^K$}
\put(324,108){\vector(-1,0){36}}
\put(315,90){$X^+_i$}
\thinlines
\put(108,108){\line(0,-1){36}}
\put(180,108){\line(0,-1){36}}
\put(252,108){\line(0,-1){36}}
\put(99,51){$\langle H \rangle$}
\put(171,51){$\langle S_i \rangle$}
\put(243,51){$\langle h_K \rangle$}
\put(162,9){{\bf Fig. 2(a)}}
\end{picture}

\begin{center}
Diagram that induces lepton masses in the model of Sec. 2.
\end{center}
\vspace{2cm}

\begin{picture}(360,216)(0,0)
\thicklines
\put(36,144){\vector(1,0){36}}
\put(36,126){$\ell^-_i$}
\put(72,144){\line(1,0){72}}
\put(99,153){$f_i$}
\put(135,126){$\ell^+_i$}
\put(180,144){\vector(-1,0){36}}
\put(180,144){\vector(1,0){36}}
\put(180,153){$f_i^S$}
\put(216,144){\line(1,0){72}}
\put(207,126){$X^-_i$}
\put(243,153){$f_i^K$}
\put(324,144){\vector(-1,0){36}}
\put(315,126){$X^+_i$}
\thinlines
\put(180,144){\oval(144,144)[b]}
\put(180,144){\line(-1,0){30}}
\put(171,103){$\langle S_i \rangle$}
\put(180,72){\line(1,-1){20}}
\put(180,72){\line(-1,-1){20}}
\put(171,76){$\lambda_K$}
\put(151,40){$\langle H \rangle$}
\put(191,40){$\langle h_K \rangle$}
\put(165,144){\oval(10,10)[tl]}
\put(165,154){\oval(10,10)[r]}
\put(165,164){\oval(10,10)[l]}
\put(165,174){\oval(10,10)[r]}
\put(180,144){\line(0,-1){30}}
\put(146,162){$\gamma$}
\put(90,108){$H$}
\put(261,108){$h_K$}
\put(162,0){{\bf Fig. 2(b)}}
\end{picture}

\begin{center}
Lepton dipole moment contribution in the model of Sec. 2.
\end{center}

\newpage

\begin{picture}(360,216)(0,0)
\thicklines
\put(36,144){\vector(1,0){36}}
\put(72,144){\line(1,0){72}}
\put(180,144){\vector(-1,0){36}}
\put(180,144){\vector(1,0){36}}
\put(177,141){$\times$}
\put(216,144){\line(1,0){72}}
\put(324,144){\vector(-1,0){36}}
\thinlines
\put(180,144){\oval(144,144)[b]}
\put(171,133){$m_a$}
\put(180,72){\line(0,-1){20}}
\put(90,108){$h_a$}
\put(261,108){$s^-_a$}
\put(171,40){$\langle H \rangle$}
\put(36,153){$L_i$}
\put(135,153){$n_a$}
\put(207,153){$n^c_a$}
\put(315,153){$\ell^+_j$}
\put(162,0){{\bf Fig. 3}}
\end{picture}

\begin{center}
Diagram that induces radiative masses for the leptons in the
model of Sec. 4.
\end{center}

\end{document}